%% file: main.tex
\documentclass[%
 reprint, superscriptaddress, nobibnotes,
 amsmath,amssymb, aps, pra
]{revtex4-2}

\usepackage{graphicx}
\usepackage{bm}
\usepackage{float}
\usepackage{hyperref}
\hypersetup{hidelinks}
\hypersetup{colorlinks = true, linkcolor=blue, urlcolor=blue,citecolor=blue}
\usepackage{physics}
\usepackage{suffix}
\usepackage{xstring}
\usepackage{siunitx}
\usepackage[dvipsnames]{xcolor}
\definecolor{darkblue}{rgb}{0.0, 0.0, 0.75}

\usepackage{color}

\usepackage[normalem]{ulem} 
\definecolor{mgreen}{RGB}{1,123,0}

\newcommand*{\refform}[1]{%
	\IfBeginWith{#1}{fig:}{Fig.~\ref{#1}}{}%
}
\newcommand*{\Rb}[1]{\ensuremath{\mathrm{^{#1}Rb}}}

\def \metre{~\mathrm{m}}

\def \m2D{\mathrm{2D}}

\begin{document}

\preprint{APS/123-QED}

\title{Detecting Phase Coherence of 2D Bose Gases via Noise Correlations}

\author{Shinichi Sunami}%
\email{shinichi.sunami@physics.ox.ac.uk}
\affiliation{Clarendon Laboratory, University of Oxford, Oxford OX1 3PU, United Kingdom}

\author{Vijay P. Singh}
\affiliation{Quantum Research Centre, Technology Innovation Institute, Abu Dhabi, UAE}

\author{Erik Rydow}%
\affiliation{Clarendon Laboratory, University of Oxford, Oxford OX1 3PU, United Kingdom}

\author{Abel Beregi}%
\affiliation{Clarendon Laboratory, University of Oxford, Oxford OX1 3PU, United Kingdom}

\author{En Chang}%
\affiliation{Clarendon Laboratory, University of Oxford, Oxford OX1 3PU, United Kingdom}

\author{Ludwig Mathey}
\affiliation{Zentrum f\"ur Optische Quantentechnologien and Institut f\"ur  Quantenphysik, Universit\"at Hamburg, 22761 Hamburg, Germany}
\affiliation{The Hamburg Centre for Ultrafast Imaging, Luruper Chaussee 149, Hamburg 22761, Germany}
\author{Christopher J. Foot}%
\affiliation{Clarendon Laboratory, University of Oxford, Oxford OX1 3PU, United Kingdom}

\date{\today}

\begin{abstract}
    We measure the noise correlations of two-dimensional (2D) Bose gases after free expansion, and use them to characterize the in-situ phase coherence across the Berezinskii-Kosterlitz-Thouless (BKT) transition. 
    The noise-correlation function features a characteristic spatial oscillatory behavior in the superfluid phase, which gives direct access to the superfluid exponent. 
    This oscillatory behavior vanishes above the BKT critical point, as we demonstrate for both single-layer and decoupled bilayer 2D Bose gases. 
    Our work establishes noise interferometry as an important general tool to probe and identify many-body states of quantum gases, extending its application to previously inaccessible correlation properties in multimode systems.
\end{abstract}

\maketitle

Noise correlations of the observables in quantum systems are characteristic features stemming from the quantum statistics of bosonic and fermionic particles and their interactions, which have numerous applications including quantum optics \cite{Brown1956}, atomic quantum technologies \cite{Aspect2019,Kaufman2018} and nuclear collisions \cite{Franz1996}.
Spatial noise correlation has proven to be a powerful method to probe the intrinsic fluctuations of many-body quantum systems.  
This was demonstrated with ultracold atoms in optical lattices \cite{Rom2006, Folling2005} and in elongated traps \cite{Dettmer2001,Manz2010,Perrin2012} through the measurement of density fluctuations after free expansion.

Density noise patterns were also observed in expanded two-dimensional (2D) Bose gases \cite{Choi2012}, 
arising from intrinsic phase fluctuations in the initially trapped cloud.  
In 2D, thermal phase fluctuations exhibit the transition from quasi-long range correlation in the superfluid phase to short-range order in the normal phase, which is the celebrated Berezinskii-Kosterlitz-Thouless (BKT) transition \cite{Berezinskii1972,Kosterlitz1973}. 
The superfluid phase is characterized by an algebraically decaying first-order correlation function $g_1(\bm{r}) = \langle \Psi^\dagger(\bm{r}) \Psi(\bm{0})\rangle/\sqrt{n(\bm{r}) n(\bm{0})} \sim |\bm{r}|^{-\eta}$, where $\Psi(\bm{r})$ is the bosonic field operator at location $\bm{r}$, $n(\bm{r}) = \langle \Psi^\dagger(\bm{r}) \Psi(\bm{r})\rangle$ is the density, and $\eta$ is the superfluid exponent. 
Theoretical studies have shown how to connect the noise correlations after the expansion to the correlation functions of the original fluctuating quantum gases \cite{Imambekov2009,Singh2014}. 
This offers a robust method to probe the in-situ spatial coherence of extended systems, a crucial quantity for investigating a wide range of many-body phases, not limited to the BKT superfluid,  
which is of great experimental interest \cite{DesbuquoisThesis, ReisThesis, SieglThesis}.



For composite low-dimensional quantum gases consisting of two constituent systems, the correlation properties of the relative and common phase modes are crucial observables characterizing the difference and sum of two fluctuating fields.
While relative modes can be probed by matter-wave interferometry \cite{Sunami2022, Gring2012}, there has not been a direct method for measuring common modes. To understand novel phases of matter predicted in coherently coupled pairs of low-dimensional systems, such as antisymmetric superfluid order in strongly coupled 2D systems \cite{Mathey2007, Benfatto2007b}, it is crucial to extend measurement capabilities to common-mode correlations.
Additionally, such a technique is essential for a complete understanding of out-of-equilibrium dynamics, where the interplay between common and relative modes exhibits unique relaxation dynamics, including near-integrable behaviour \cite{Gring2012} and universal scaling \cite{Sunami2023}.



In this Letter, we report the measurement of the noise correlations of expanded single and decoupled bilayer 2D Bose gases.
For the bilayer, this provides information about common-mode phase fluctuations.
Using analytical models we determine the superfluid exponent $\eta$ for a wide range of the phase-space densities across the BKT transition. 
The results are consistent for single and decoupled bilayer systems, as expected in the absence of interlayer coupling. 
We benchmark these results with those determined from the measurement of the relative phase of our bilayer system using local matter-wave interferometry \cite{Sunami2022}.

\begin{figure*}[ht]
    \includegraphics[width=0.99\textwidth]{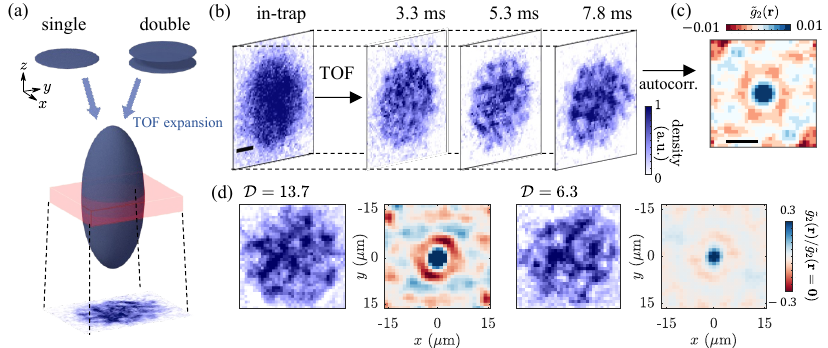}
        \caption{\label{fig:main} Noise interferometry of 2D Bose gases. 
        (a) Sketch of the noise detection. 
        We trap single- or double-layer ultracold 2D Bose gases  with tight vertical confinement, 
        and release them for a time-of-flight (TOF) expansion of duration $t$.
        The tight vertical confinement results in a fast vertical expansion of the cloud without significant expansion in the radial direction. 
        We record a slice of the density distribution using absorption imaging with a selective optical repumping technique (red transparent sheet), which is necessary since the depth of field of the high-resolution imaging apparatus is smaller than the vertical extent of the expanded cloud. 
        (b) Single-shot images of the density distribution of trapped single-layer clouds with no expansion (in-situ) and at $t = 3.3$, 5.3 and 7.8 $ \SI{}{ \milli \second}$.
        (c) The noise correlation function $\tilde{g}_2(\bm{r})$, obtained by calculating the density-density correlations and averaging over experimental repeats; see text.
        A characteristic ring structure emerges at distance $|\bm{r}| \sim 7 \SI{}{\micro \meter}$ due to density modulations that are present in the expanded cloud at $t= \SI{7.8}{ \milli \second}$.
        The black bars in panels (b) and (c) denote $\SI{10}{\micro \metre}$.
        (d) Typical single-shot density distribution and $\tilde{g}_2(\bm{r})$ shown after $t=\SI{5.3}{ \milli \second}$ for decoupled bilayer systems with phase-space densities of $\mathcal{D} = 13.7$ and $6.3$.
    }
\end{figure*}

We prepare 2D degenerate Bose gases of \Rb{87} atoms in a cylindrically-symmetric trap having strong confinement along the $z$ direction.
In this work, we create single-well or double-well vertical confinement using either a single-RF or multiple-RF dressing technique as described in \cite{Perrin2017,Luksch2019,Barker2020jphysb,Barker2020,Sunami2022}.
Near the minimum of each well, in the single or double-well potential, the vertical trap frequencies are $\omega_z/2\pi  = \SI{1.2}{kHz}$ giving a dimensionless 2D interaction strength $\tilde{g} = \sqrt{8\pi} a_s/\ell_0=0.08$, where $a_s$ is the 3D scattering length, $\ell_0=\sqrt{\hbar/(m\omega_z)}$ is the harmonic oscillator length along $z$ for an atom of mass $m$ and 
$\hbar$ is the reduced Planck constant. 
We impose additional optical trapping in the horizontal plane using a ring-shaped strong off-resonant laser beam at 532 nm to realize near-homogeneous 2D systems \cite{supp}.
We load atoms into the trap at a temperature $T=\SI{50}{\nano \kelvin}$, set by forced evaporation. 
For double-well potentials we equalize the population $N$ in each layer by maximizing the matter-wave interference contrast \cite{Barker2020, Sunami2022}. 
A large inter-well separation of $\SI{7}{\micro \meter}$ gives a decoupled bilayer system.
We vary $N$ to cover a broad range of the phase-space density $\mathcal{D} = n h^2 /(2\pi m k_B T)$ between $5$ and $24$, where $n$ is the average 2D density, and $k_B$ is the Boltzmann constant.
For all parameters employed, the quasi-2D conditions $\hbar \omega_z > k_B T$ and $\hbar \omega_z > \mu$ are satisfied, where $\mu$ is the chemical potential \cite{supp}. 

The detection scheme is illustrated in Fig.\ \ref{fig:main}(a). 
The trap is abruptly turned off to release the 2D gas for a time-of-flight (TOF) expansion of duration $t$ between 3.3 and 7.8 ms \cite{trap_pos,Sunami2022}, starting with either one or two clouds.
Once released, the clouds expand ballistically along the $z$ direction with negligible expansion in the radial direction.
We image a slice of the expanded clouds with thickness $L_z = \SI{10}{\micro\metre}$ along the $z$ direction which is much smaller than the extent of the expanded cloud in this direction, as illustrated in Fig.\ \ref{fig:main}(a).
This ensures that the imaged sample is thin compared to the depth of field of the high-resolution imaging system \cite{supp}; further, we carefully focus the imaging system using the density-noise patterns as described in the Suppl.~Mat.~\cite{supp}. 
This minimizes the systematic effect of imperfect imaging on the density pattern measurements \cite{Choi2012,Langen2013}.
Examples of the density distribution after several values of the TOF duration $t$, including the in-situ density distribution at $t=0$, are shown in Fig.\ \ref{fig:main}(b).
The system is deep in the superfluid regime with peak PSD of $\mathcal{D} = 24$ \cite{Sunami2022, supp}.
Compared to the in-situ density distribution, the expanded cloud shows the formation of density modulations on a length scale growing with $t$ because in self-interference of the cloud the interference occurs at distances that depend on $t$ \cite{Imambekov2009,Choi2012,Singh2014}. 
As a result, initial phase fluctuations transform into density modulations that we analyze. 
We autocorrelate the normalized density distribution in each cloud and then average the density-density correlations over at least 20 repetitions.
Thus, we obtain the noise correlation function \cite{supp, Singh2014}
\begin{equation}\label{eq:g2_def_main}
    g_2(\bm{r}, t) = \frac{  \langle \Psi^{\dagger}(\bm{r}, t) \Psi^{\dagger}(\bm{0}, t) \Psi(\bm{r}, t)\Psi(\bm{0}, t) \rangle  }{n(\bm{r}, t) n(\bm{0}, t)}.
\end{equation}
In Fig.\ \ref{fig:main}(c), we show $\tilde{g}_2 (\bm{r}) \equiv g_2(\bm{r})-1$ at $t = \SI{7.8}{\milli \second}$, 
which features a characteristic minimum at distance $r\sim \SI{7}{\micro \meter}$ as predicted from the theoretical model \cite{Singh2014}.
We repeat these measurements for a decoupled bilayer system. 
Examples of the single-shot density distribution and averaged noise correlation function $\tilde{g}_2 (\bm{r}, t)$, taken with $t = \SI{5.3}{\milli \second}$ for the systems at $\mathcal{D}=13.7$ and $6.3$, are shown in Fig.\ \ref{fig:main}(d). 
The existence of the characteristic minimum at high $\mathcal{D}$ is a feature of the superfluid phase, which is suppressed for the normal phase at lower $\mathcal{D}$ \cite{Singh2014}.

To infer the properties of the system we analyze how $g_2(\bm{r}, t)$ depends on the expansion time and the in-situ phase fluctuations. 
Analysis of free expansion of a single-layer 2D quasicondensate with phase fluctuations yields $g_2(\bm{r}, t)$ as a function of expansion time $t$ \cite{Imambekov2009, Singh2014, supp}
\begin{eqnarray}\label{eq:g2_g1}
	g_2(\bm{r}, t ) = \frac{1}{(2\pi)^2}\int d^2 \bm{q} \int d^2 \bm{R} \cos(\bm{q}\cdot \bm{r})\cos(\bm{q}\cdot \bm{R}) \nonumber \\
	\times \frac{\mathcal{F}(\bm{q}_t)^2 \mathcal{F}(\bm{R})^2}{\mathcal{F}(\bm{R}-\bm{q}_t)\mathcal{F}(\bm{R}+\bm{q}_t)},
\end{eqnarray}
where $\bm{q}_t = \hbar \bm{q} t/m$ with $\bm{q}$ being the wave vector and the functions $\mathcal{F}$ are the first-order correlation functions i.e. $\mathcal{F}(\bm{r}) \sim g_1(\bm{r})$. 
This predicts distinctly different behavior for $g_2(\bm{r}, t )$ for the algebraic and exponential forms of $g_1(\bm{r})$, which we utilize to probe the BKT transition \cite{Singh2014, supp}.   

\begin{figure}[t]
    \includegraphics[width=0.99	\linewidth]{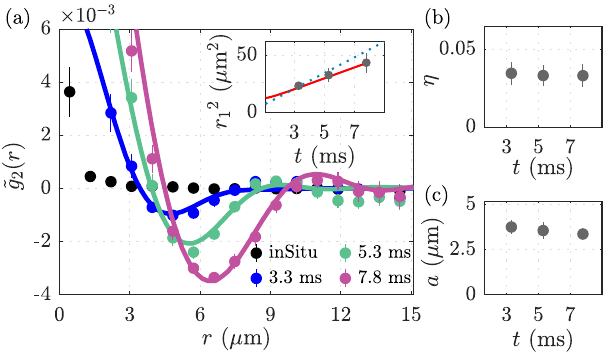}
        \caption{\label{fig:tof} Expansion scaling behavior.
        (a) Radially averaged correlation function $\tilde{g}_2(r,t)$ is shown for trapped single-layer clouds with no expansion (in-situ) and after $t =3.3$, 5.3 and 7.8 $ \SI{}{ \milli \second}$. 
        Error bars are the standard error.
        Solid lines are the best fits based on Eq.\ \eqref{eq:g2_g1} for the superfluid phase, 
        from which we determine the superfluid exponent $\eta$ and the short-range cutoff $a$.  
        Inset shows the squared locations $r_1^2$ of the first minimum of the correlation function (dots), 
        together with the mean-field prediction $r_1 {}^2 \propto t$ (dashed line). 
        The solid line shows the analytical prediction of the minimum position in the superfluid phase based on Eq.\ \eqref{eq:g2_g1}, 
        including the effect of finite imaging resolution encoded in the short-range cutoff.  
        (b, c) Extracted values of $\eta$ and $a$.
	   }
\end{figure}

In Fig.\ \ref{fig:tof}(a), we show the measurements of $\tilde{g}_2(r,t)$ for a single-layer cloud deep in the superfluid regime, with $\mathcal{D}=24$, for several expansion times $t$.
We shift the locations of the trapped cloud with a magnetic bias field along $z$ for each $t$, without changing any system parameters such as trap geometry, temperature and atom number, such that the expanded cloud is at the focus of the imaging apparatus after expansion time $t$.
For the quantitative analysis, we determine the radially averaged $g_2(r,t)$ as a function of the radial coordinate $r$. 
Oscillatory behavior emerges for expanded clouds, which is absent for in-trap measurements.
We fit the measurements at finite $t$ with the predictions of Eq.\ \eqref{eq:g2_g1} for the superfluid phase  to determine the superfluid exponent $\eta$ and the short-range cutoff $a$. 
The superfluid model describes our measurements well, with consistent values of $\eta \simeq 0.04$ and $a \simeq \SI{3}{\micro \meter}$ for all three expansion times as expected for the same in-situ system parameters, see Figs.\ \ref{fig:tof} (b,c). 
To further corroborate the expansion scaling, we compare the locations $r_1$ of the first minimum of $\tilde{g}_2(r, t)$ with the mean-field prediction $r_1 {}^2=3\pi \hbar t/m$ and the predictions of the superfluid model, see inset of Fig.\ \ref{fig:tof} (a).

\begin{figure*}[t]
    \includegraphics[width=0.99	\textwidth]{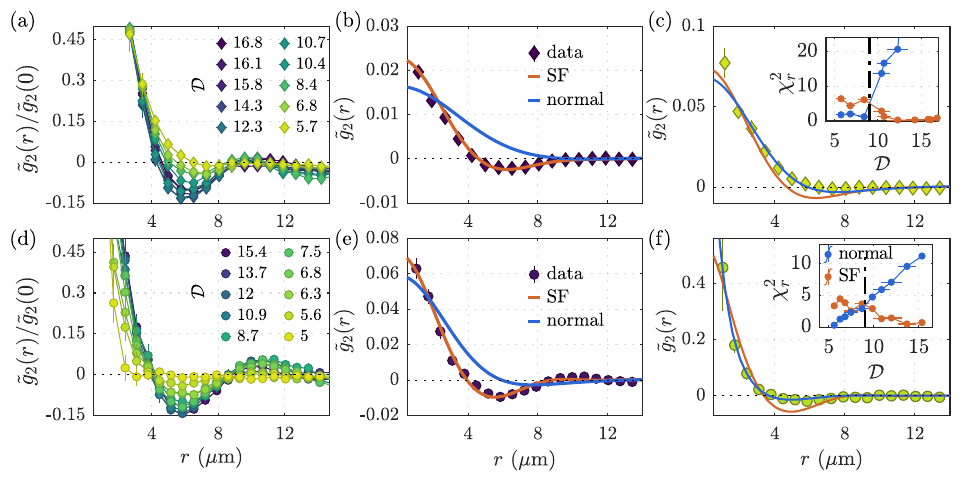} 
    \caption{\label{fig:g2} Noise correlations of single layer and bilayer 2D Bose gases across the BKT transition.
    (a) $\tilde{g}_2(r)$, scaled by the value at origin $\tilde{g}_2(0)$, for $t=5.3$ ms for single-layer clouds having phase-space densities between $\mathcal{D}=5.7$ and $16.8$.
    Lines are guides to the eye and error bars are the standard error.
    (b) $\tilde{g}_2(r)$ at $\mathcal{D} = 16.8$, plotted together with best fits of superfluid (SF, orange) and normal phase (blue) based on Eq.\ \eqref{eq:g2_g1}.
    (c) Measurement at $\mathcal{D} = 5.7$ is well described by the prediction for the normal phase. 
    Inset shows the reduced $\chi^2$ statistics of the superfluid and normal-phase fitting models at varying $\mathcal{D}$ (values for the normal phase at high $\mathcal{D}$ with $\chi_r^2 > 20$ are not shown).
    The crossover from superfluid to normal phase is identified by the point (vertical dashed line) beyond which $\tilde{g}_2(r)$ is better described by the normal-phase model at small $\mathcal{D}$. This yields the crossover at $\mathcal{D}_c = 9.1(9)$.
    (d) Measurements after $t=5.3$ ms for bilayer clouds with phase-space densities between $\mathcal{D}=5$ and $15.4$. 
    (e) $\tilde{g}_2(r)$ and their fitted curves at $\mathcal{D}=15.4$.
    (f) The measurement at $\mathcal{D}=5.6$ is better described by the normal-phase model, similarly to the single-layer case at small $\mathcal{D}$ in panel (c). 
    Inset showing the $\chi_r^2$ errors of the fitting model, whose intersection gives the BKT crossover at $\mathcal{D}_c = 9.0(6)$ (vertical dash-dotted line). 
    }
\end{figure*}

Having verified the expansion scaling behavior, we now turn our attention to the measurements across the BKT critical point, both with single and uncoupled bilayer 2D clouds.
In Fig.\ \ref{fig:g2}(a), we show the measurements of $\tilde{g}_2(r)$ for single-layer clouds at varying $\mathcal{D}$. 
As $\mathcal{D}$ is lowered, the oscillatory feature of the noise correlations, characteristic of the superfluid phase, vanishes indicating a crossover to the normal phase.
To verify this we fit the measurements with the predictions of Eq.\ \eqref{eq:g2_g1} calculated for both the superfluid and normal phase as shown in Figs.\ \ref{fig:g2} (b-c) for both large and small $\mathcal{D}$;
for $\mathcal{D} = 16.8$ the prediction for the superfluid phase gives a better fit, whereas for $\mathcal{D} = 5.7$ it is the normal phase.
We determine the BKT crossover point as being the value of $\mathcal{D}$ at which the exponential model overtakes the superfluid model, see inset of Fig.\ \ref{fig:g2}(c). 
This gives the crossover point at $\mathcal{D}_c = 9.1(9)$, which is in close agreement with the theoretical prediction $\mathcal{D}_{c,th}=8.6$ \cite{Prokofev2001}.  

In Fig.\ \ref{fig:g2}(d) we show the measurements of our bilayer systems at varying $\mathcal{D}$. 
Similar to the single-layer case, the noise correlations are oscillatory at large $\mathcal{D}$, however, there is a fast short-distance fall off of correlations in the normal regime at small $\mathcal{D}$. 
This is the characteristics of density-noise functions for bilayer systems, as we discuss in the following.
%
For a bilayer 2D Bose gas described by the field operators $\Psi_i(\bm{r}, t)$ for layers $i=1,2$, the density-noise pattern measured along the $z$ axis arises predominantly from by the common-mode fluctuations with correlation function 
$ \mathcal{F}_{\mathrm{com}}(\bm{r})^2 \simeq  \langle \Psi_1^{\dagger}(\bm{r})\Psi^{\dagger}_2(\bm{r}) \Psi_1(\bm{0})\Psi_2(\bm{0}) \rangle/n^2$, 
i.e. the noise correlation function of the expanded bilayer, which is defined by replacing the field operators in Eq.\ \eqref{eq:g2_def_main} by $\Psi(\bm{r}, t) = (\Psi_1(\bm{r}, t)+\Psi_2(\bm{r}, t))/\sqrt{2}$. 
For a decoupled bilayer the noise correlation function is \cite{supp}
\begin{eqnarray}\label{eq:g2_g1_com} 
	g_2(\bm{r}, t ) \approx  \dfrac{1}{(2\pi)^2} \int d^2 \bm{q} \int d^2 \bm{R} \cos(\bm{q}\cdot \bm{r})\cos(\bm{q}\cdot \bm{R}) \nonumber \\
	\times \dfrac{\mathcal{F}_{\mathrm{com}}(\bm{q}_t)^2 \mathcal{F}_{\mathrm{com}}(\bm{R})^2}{\mathcal{F}_{\mathrm{com}}(\bm{R}-\bm{q}_t)\mathcal{F}_{\mathrm{com}}(\bm{R}+\bm{q}_t)} \mathcal{F}_{\mathrm{rel}}(\bm{q}_t)^2,
\end{eqnarray}
where finite mixing of relative phase modes with correlation function $\mathcal{F}_{\mathrm{rel}}(\bm{r})^2 \simeq  \langle \Psi_1^{\dagger}(\bm{r})\Psi_2(\bm{r}) \Psi_2^{\dagger}(\bm{0})\Psi_1(\bm{0}) \rangle/n^2$ affects $g_2(\bm{r}, t )$ in the normal regime, resulting in a fast decay of $|g_2|$ at short distances, see Fig.\ S1 in \cite{supp}.
We calculate the predictions of Eq.\ \eqref{eq:g2_g1_com} for the superfluid and normal regimes and use them to fit our bilayer noise correlations to determine the exponent $\eta$ and the BKT crossover, see Figs.\ \ref{fig:g2} (d-f). 
The BKT crossover occurs at $\mathcal{D}_c = 9.0(6)$, which is in agreement with the results of the single layer behavior and the theoretical prediction \cite{Prokofev2001}. 
Combined with the relative-mode phase correlation measurements using matter-wave homodyning \cite{Hadzibabic2006,Sunami2022}, this method gives a complete characterization of phase fluctuations of bilayer systems. 
The relative and common modes have the same statistics in the absence of interlayer coupling, which we utilize to benchmark the results of the noise measurements \cite{Sunami2022, supp}, as we discuss below.   

\begin{figure}[ht]
    \includegraphics[width=0.99	\linewidth]{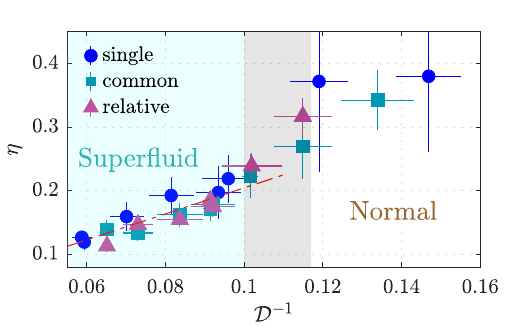}
    \caption{\label{fig:eta} The BKT crossover in 2D Bose gases.  
    The algebraic exponent $\eta$ is plotted as a function of the inverse phase-space density 
    $\mathcal{D}^{-1}$. 
    Results for a single layer (circles) and bilayer (squares) are compared against the measurements (triangles) of the relative-phase correlations \cite{supp}. 
    Red dash-dotted line is the Nelson-Kosterlitz relation $\eta \propto \mathcal{D}^{-1}$, expected in the superfluid regime \cite{Nelson1977}.
    The crossover region (black shaded area) is at $\mathcal{D}^{-1}= $ 0.11(1), corresponding to a critical algebraic exponent $\eta_c \sim 0.2$ as expected for a finite-sized system \cite{Sunami2022}.
    Above the critical point, values of $\eta$ have increased uncertainty as the in-situ phase correlations become short ranged.
    }
\end{figure}

In Fig.\ \ref{fig:eta} we consolidate the results for single and bilayer 2D clouds. 
The measured value of $\eta$ scales linearly with the inverse phase-space density $\mathcal{D}^{-1}$ in the superfluid regime, in accordance with the Nelson-Kosterlitz relation $\eta \propto \mathcal{D}^{-1}$ \cite{Nelson1977,Sunami2022}.
There is a deviation from this linear behavior when the system approaches the crossover regime at $\mathcal{D}^{-1}= $ 0.11(1), as predicted \cite{Sunami2022}: in the normal phase above the broad crossover regime $\eta$ shows a faster increase within the measurement uncertainty.
The determination of $\eta$ is less accurate in the normal regime since the correlations are short ranged and better described by exponential scaling.  
We also show values of $\eta$ obtained from the relative-phase correlation measurements \cite{supp,Sunami2022}, 
which agree with the results of the noise measurements. 

In conclusion, we have demonstrated a novel experimental technique to probe phase fluctuations in 2D Bose gases through the correlation analysis of density noise patterns appearing after short time-of-flight (TOF) expansions.
The location $r_1$ of first minimum of the measured noise correlation function $g_2(r,t)$ scales with the expansion time $t$ as $r_1^2 \propto t$.  
This confirms the expansion scaling and demonstrates that the determined values of the superfluid exponent and the BKT critical point are independent of the expansion time.
Furthermore, the noise interferometry of expanded bilayer systems provides a direct access to the common-mode phase correlations, which we benchmarked by exploiting the fact that, in the absence of interlayer coupling, they exhibit the same behavior as the the relative-mode correlation function. 

The spatial noise interferometry method presented in this Letter, when used in conjunction with the direct matter-wave interferometric probe of the relative phase mode \cite{Sunami2022}, contributes to the development of a comprehensive understanding of novel many-body phases of matter. 
For instance, superfluid phases arising from coherent coupling of bilayer 2D quantum systems, where relative and common modes are expected to exhibit qualitatively distinct features, represent a class of novel composite orders \cite{Mathey2007, Benfatto2007b, Song2022, Eto2018, Furutani2023, Bighin2019} that can only be identified and elucidated through the combined measurement approach demonstrated in this Letter.


\textit{Note added.} After submitting our manuscript, we became aware of a relevant study that uses density ripple information to determine the total phase of 1D Bose gases \cite{Murtadho2024}.

\begin{acknowledgements}
    This work was supported by the EPSRC Grant Reference EP/X024601/1. 
	A. B. and E. R. thank the EPSRC for doctoral training funding.  
	L. M. acknowledges funding by the Deutsche Forschungsgemeinschaft (DFG) in the framework of SFB 925 – project ID 170620586, the excellence cluster `Advanced Imaging of Matter’ - EXC 2056 - project ID 390715994, and by the Hamburg Quantum Computing (HQC) initiative.
	The project is co-financed by ERDF of the European Union and by `Fonds of the Hamburg Ministry of Science, Research, Equalities and Districts (BWFGB)'.
\end{acknowledgements}

\input{refs.bbl}

\setcounter{equation}{0}
\setcounter{figure}{0}
\setcounter{table}{0}
\renewcommand{\theequation}{S\arabic{equation}}
\renewcommand{\thefigure}{S\arabic{figure}}

\onecolumngrid
\newpage
\section*{Supplemental material}

\subsection{Density-noise correlation function of expanded bilayer 2D Bose gases}\label{sec:theory}

In this section, we analyze the density-noise correlation function of expanded bilayer systems to obtain the expression based on in situ correlation properties, following a similar derivation as for the single-layer systems in Ref.~\cite{Singh2014}.
The bilayer system consists of two layers $i=1,2$ initially with separation $d$, where $d=\SI{7}{\micro\meter}$ for the experiments reported in the main text.
We express the two layers using bosonic field operators $\Psi_{i}(\bm{r})$ and take the initial 3D wavefunction to be of the form
\begin{equation}
	\hat{\bm{\Psi}}_0(\bm{r}_{2D},z;t)= [\hat{\bm{\Psi}}_1(\bm{r})\delta(d/2-z) + \hat{\bm{\Psi}}_2(\bm{r})\delta(d/2+z)]/\sqrt{2},
	\label{eq:separationofvariables}
\end{equation}
and analyze its expansion dynamics.
Free expansion of the bosonic field operator is given by 
\begin{equation}
    \hat{\bm{\Psi}}(\bm{r}, t) =  \int d^3 \bm{r}' G_3(\bm{r}-\bm{r}',t)\hat{\bm{\Psi}}(\bm{r}', 0),
    \label{eq:bosonfieldoperator}
    \end{equation}
\noindent where the 3D Green's function is 
\begin{equation}
    G_3(\bm{r}-\bm{r}',t) =  G_1(x-x',t)G_1(y-y',t)G_1(z-z',t),
    \label{eq:3dgreens}
\end{equation}
and 
\begin{equation}
    G_1(r,t)= \sqrt{\frac{m}{2 \pi i \hbar t}} e^{\frac{i m}{2 \hbar t} r^2}.
    \label{eq:3dgreen}
\end{equation}
Combining the prefactors into $A=\frac{m}{2 \pi \hbar t}$, the density distribution after the TOF expansion is
\begin{equation}
    \hat{n}(\bm{r},t) = A^3\int d^3 \bm{v} \int d^3 \bm{w} \ e^{\frac{-im}{2 \hbar t}(\bm{r}-\bm{v})^2}\hat{\bm{\Psi}}^{\dag}(\bm{v}, 0) e^{\frac{im}{2 \hbar t}(\bm{r}-\bm{w})^2}\hat{\bm{\Psi}}(\bm{w}, 0).
\end{equation}
Following the procedure in Ref.\ \cite{Singh2014}, denoting the spatial coordinates in the $xy$-plane by the subscript $\parallel$ and coordinates perpendicular to it (i.e. in the $z$ direction) by the subscript $\bot$, the density-density correlation function is
\begin{equation}
	\begin{aligned}
		g_2(\bm{r}_1, \bm{r}_2, t)  = &  \frac{A^4}{4 n_0^2} \int d^2 \bm{r} \int d^2 \bm{r}' \int d s_{\bot}  \int d S_{\bot} \int d u_{\bot} \int d U_{\bot} \\
		&\times e^{\frac{-im}{4 \hbar t} \left [\bm{r}^2 - \bm{r}'^2 - 2 \bm{r}_{\parallel 12} \cdot (\bm{r}- \bm{r}') \right ] }  
		 \times e^{\frac{-im}{\hbar t} \left [ (S_{\bot}-\frac{r_{\bot 12}}{2})s_{\bot} + (U_{\bot}+\frac{r_{\bot 12}}{2})u_{\bot} - R_{\bot 12}(s_{\bot}+ u_{\bot}) \right ]} \\
		& \times \biggl< \hat{\bm{\Psi}}_0^{\dag}(\frac{\bm{r}}{2}, S_{\bot}+\frac{s_{\bot}}{2}) \hat{\bm{\Psi}}_0^{\dag}(-\frac{\bm{r}}{2}, U_{\bot}+\frac{u_{\bot}}{2}) 
		\hat{\bm{\Psi}}_0(\frac{\bm{r}'}{2}, S_{\bot}-\frac{s_{\bot}}{2})   \hat{\bm{\Psi}}_0(-\frac{\bm{r}'}{2}, U_{\bot}-\frac{u_{\bot}}{2}) \biggr>,
	\end{aligned}
	\label{eq:g2_2dsimplified}
\end{equation}

\noindent where we use center of mass coordinates, $\bm{r}_{12}=\bm{r}_1 - \bm{r}_2$ and $\bm{R}_{12}=(\bm{r}_1 + \bm{r}_2)/2$.
With integration along the $z$ direction with sufficient integration width, equivalent to the imaging procedure in the experiment reported in the main text \cite{integrate_out}, we obtain in terms of 2D coordinates,
\begin{equation}
	\begin{aligned}
		\Tilde{g}_2(\bm{r}_{1, 2D}, \bm{r}_{2, 2D})  = \int d r_{1 z} \int d r_{2 z}  g_2(\bm{r}_{1, 3D}, \bm{r}_{2, 3D}),
	\end{aligned}
	\label{eq:g2_tilde}
\end{equation}
where the integration over $r_{1 z}$, $r_{2 z}$ (or $R_{\bot 12}$ and $r_{\bot 12}$), gives delta functions for $u_{\bot}$ and $s_{\bot}$. 
The delta functions result in $s_{\bot}= - u_{\bot}$ and $s_{\bot}=u_{\bot}=0$ and thus modify the prefactor, such that
\begin{equation}
	\begin{aligned}
		\Tilde{g}_2(\bm{r}_{1, 2D}, \bm{r}_{2, 2D})   =   \frac{A^2}{4 n_0^2} \int d^2 \bm{r} & \int d^2 \bm{r}' \int d S_{\bot} \int d U_{\bot} 
		e^{\frac{-im}{4 \hbar t} \left [\bm{r}^2 - \bm{r}'^2 - 2 \bm{r}_{\parallel 12} \cdot (\bm{r}- \bm{r}') \right ] }  \\ 
		& \times \biggl< \hat{\bm{\Psi}}_0^{\dag}(\frac{\bm{r}}{2}, S_{\bot}) \hat{\bm{\Psi}}_0^{\dag}(-\frac{\bm{r}}{2}, U_{\bot}) \hat{\bm{\Psi}}_0(\frac{\bm{r}'}{2}, S_{\bot})   \hat{\bm{\Psi}}_0(-\frac{\bm{r}'}{2}, U_{\bot}) \biggr>.
	\end{aligned}
	\label{eq:g2_2dafterdelta_simlified}
\end{equation}
We now reintroduce the original $z$ spatial variables, $S_{\bot} = (z_1 + z_2)/2$, $s_{\bot} = z_1 - z_2$ and $U_{\bot} = (z_3 + z_4)/2$, $u_{\bot} = z_3 - z_4$. 
With the delta functions for  $s_{\bot}$ and $ u_{\bot}$ described above we have $S_{\bot} = z_1$, $U_{\bot} = z_3$, so that
\begin{equation}
	\begin{aligned}
		\Tilde{g}_2(\bm{r}_{1, 2D}, \bm{r}_{2, 2D})   =  \frac{A^2}{4 n_0^2} \int d^2 \bm{r} & \int d^2 \bm{r}' \int d z_1 \int d z_3  e^{\frac{-im}{4 \hbar t} \left [\bm{r}^2 - \bm{r}'^2 - 2 \bm{r}_{\parallel 12} \cdot (\bm{r}- \bm{r}') \right ] }  \\ &
		\times \biggl< \hat{\bm{\Psi}}_0^{\dag}(\frac{\bm{r}}{2}, z_1) \hat{\bm{\Psi}}_0^{\dag}(-\frac{\bm{r}}{2}, z_3) \hat{\bm{\Psi}}_0(\frac{\bm{r}'}{2}, z_1)   \hat{\bm{\Psi}}_0(-\frac{\bm{r}'}{2}, z_3) \biggr>.
	\end{aligned}
	\label{eq:g2_2dafterdelta_simlified2}
\end{equation}
\begin{figure}[t]
	\includegraphics[width=0.6	\linewidth]{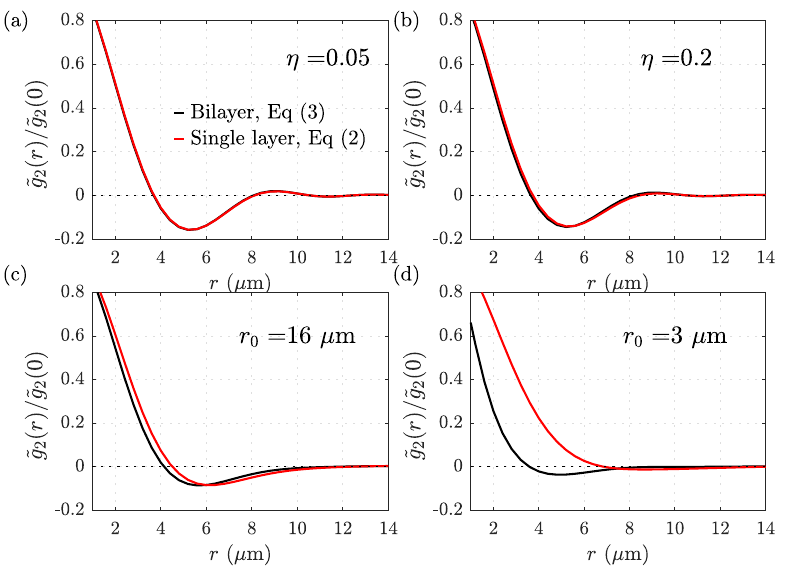}
	\caption{\label{fig:doublewelltheory} Theoretical model of bilayer density-noise correlation functions with (red) and without (black) considering the mixing of relative phase effect, for (a,b) superfluid regime, and (c,d) normal regime.
        The theoretical curves are obtained by the numerical integration of Eqs.\ (2) and (3) in a similar manner to Ref.~\cite{Singh2014}, where for both common-mode and relative-mode correlation function $\mathcal{F}_{\mathrm{com}}$ and $\mathcal{F}_{\mathrm{rel}}$ are taken to be the first-order correlation function of a single layer system, as applicable for a decoupled bilayer.
		}
\end{figure}
\noindent To simplify notation we introduce 
\begin{equation}
	\rho_{ijkl}(\bm{r}, \bm{r}') = \biggl< \hat{\bm{\Psi}}_i^{\dag}(\frac{\bm{r}}{2})\hat{\bm{\Psi}}_j^{\dag}(-\frac{\bm{r}}{2})\hat{\bm{\Psi}}_k(\frac{\bm{r}'}{2})\hat{\bm{\Psi}}_l(-\frac{\bm{r}'}{2}) \biggr>.
	\label{eq:rho_notation}
\end{equation}
Substituting Eq.~\ref{eq:separationofvariables} into Eq.~\ref{eq:g2_2dafterdelta_simlified2} and evaluating the $z_1$ and $z_2$ integrals,
\begin{equation}
	\begin{aligned}
		\Tilde{g}_2(\bm{r}_{1, 2D}, \bm{r}_{2, 2D})   =   \frac{A^2}{4 n_0^2} & \int d^2 \bm{r} \int d^2 \bm{r}' e^{\frac{-im}{4 \hbar t} \left [\bm{r}^2 - \bm{r}'^2 - 2 \bm{r}_{\parallel 12} \cdot (\bm{r}- \bm{r}') \right ] }  \\ &
		\times \frac{1}{4}\biggl[ \rho_{1111}(\bm{r}, \bm{r}') + \rho_{2121}(\bm{r}, \bm{r}')  + \rho_{1212}(\bm{r}, \bm{r}') + \rho_{2222}(\bm{r}, \bm{r}')\biggr].
	\end{aligned}
	\label{eq:g2_2dafter2dwf}
\end{equation}
Using the density-phase representation of the Bose field in the trap, $\Psi(\bm{r}) \simeq \sqrt{n} \exp(i\phi(\bm{r}))$, applicable for our case of quasicondensates \cite{Posazhennikova2006}, Eq.\ \eqref{eq:rho_notation} can be represented by the phase fluctuation as
$\rho_{ijkl}(\bm{r}, \bm{r}') \sim  n_0^2 \left <  e^{i[\theta_i(\bm{r}/2)+ \theta_j(-\bm{r}/2) - \theta_k(\bm{r}'/2) - \theta_l(-\bm{r}'/2)]}\right >$ where $\theta(\bm{r})=(\phi_1(\bm{r})-\phi_2(\bm{r}))/2$, $\varphi(\bm{r})=(\phi_1(\bm{r})+\phi_2(\bm{r}))/2$ are the antisymmetric and symmetric phase modes.
Writing the complex exponentials using cosines, the density-density correlation function of expanded cloud is
\begin{equation}
	\begin{aligned}
		\Tilde{g}_2(\bm{r}_{1, 2D}, \bm{r}_{2, 2D})   =  \frac{A^2}{4 n_0^2} & \int d^2 \bm{r} \int d^2 \bm{r}' 
		   e^{\frac{-im}{4 \hbar t} \left [\bm{r}^2 - \bm{r}'^2 - 2 \bm{r}_{\parallel 12} \cdot (\bm{r}- \bm{r}') \right ] } \\ &
		\times \rho_{\varphi}(\bm{r}, \bm{r}')  \biggl< \cos (\Delta \theta(\bm{r}, \bm{r}') ) \cos (\Delta \theta(-\bm{r}, -\bm{r}') ) \biggr>,
	\end{aligned}
	\label{eq:g2_2dafterdelta}
\end{equation}
where $\rho_{\varphi}(\bm{r}, \bm{r}') = n_0^2 \left <  e^{i[\varphi(\bm{r}/2)+ \varphi(-\bm{r}/2) - \varphi(\bm{r}'/2) - \varphi(-\bm{r}'/2)]}\right > $ expresses the common-mode phase correlation which dominantly contributes to the measured density noise correlation;
the final term containing $\Delta \theta(\bm{r}, \bm{r}') = \theta(\bm{r}/2) - \theta(\bm{r}'/2)$ encapsulates the effect of finite mixing of relative phase on $\Tilde{g}_2$. 
Introducing the shorthand notation $\Delta \theta(\pm) = \Delta \theta(\pm \bm{r}, \pm \bm{r}') $, this term can be expressed as exponentials
\begin{equation}
    \biggl< \cos (\Delta \theta(+) ) \cos (\Delta \theta(-) ) \biggr> = 
    \frac{1}{4} \biggl< e^{i[\Delta \theta(+)+ \Delta \theta(-)]}  + e^{i[\Delta \theta(+)- \Delta \theta(-)]} + e^{i[-\Delta \theta(+)+ \Delta \theta(-)]} + e^{i[-\Delta \theta(+)- \Delta \theta(-)]}\biggr>,
\label{eq:common_phase_1}
\end{equation}
assuming Gaussian phase fluctuations $\left < e^{i[\theta(r)+\theta(r')]}\right> = e^{-\frac{1}{2}\left < [\theta(r)+\theta(r')]^2 \right>}$. 
Defining the relative-phase correlation function $\mathcal{F}_{\mathrm{rel}}(\bm{r}-\bm{r}') \equiv \left < e^{i[\theta(\bm{r})-\theta(\bm{r}')] }\right> \simeq  \langle \hat{\Psi}_1^{\dagger}(\bm{r}, 0)\hat{\Psi}_2(\bm{r}, 0) \hat{\Psi}_2^{\dagger}(\bm{0}, 0)\hat{\Psi}_1(\bm{0}, 0) \rangle/n_0^2$ and expanding the exponents for each term in Eq.~\ref{eq:common_phase_1},
\begin{equation}
	\begin{aligned}
		\biggl< \cos (\Delta \theta(+) ) \cos (\Delta \theta(-) ) \biggr> \approx \mathcal{F}_{\mathrm{rel}}([\bm{r}- \bm{r}']/2)^2 ,
	\end{aligned}
	\label{eq:coscosg2}
\end{equation}
to linear order in $\left < \Delta \theta(-)\Delta \theta(+) \right >$. Thus the density correlation function is
\begin{equation}
		\Tilde{g}_2(\bm{r}_{1, 2D}, \bm{r}_{2, 2D})   \approx \frac{A^2}{4 n_0^2} \int d^2 \bm{r} \int d^2 \bm{r}' e^{\frac{-im}{4 \hbar t} \left [\bm{r}^2 - \bm{r}'^2 - 2 \bm{r}_{\parallel 12} \cdot (\bm{r}- \bm{r}') \right ] } \rho_{\varphi}(\bm{r}, \bm{r}') \mathcal{F}_{\mathrm{rel}}([\bm{r}- \bm{r}']/2)^2  .
	\label{eq:g2_final}
\end{equation}

\noindent Following the procedure for the case of a single layer~\cite{Singh2014} we perform a change of variables,  $ \bm{q} =  (m / 2 \hbar t)(\bm{r} - \bm{r}') $, $\bm{R}=(\bm{r} + \bm{r}')/2$, effectively resulting in a Fourier transform. 
Rewriting the result using $\mathcal{F}_{\mathrm{com}}(\bm{r})$ and $\bm{q}_t$ defined as in the main text, gives
\begin{equation}
		\Tilde{g}_2(\bm{r}, t)   \approx \frac{1}{(2\pi)^2} \int d^2 \bm{q} \int d^2 \bm{R} \cos(\bm{q}\cdot \bm{r})\cos(\bm{q}\cdot \bm{R})  \frac{\mathcal{F}_{\mathrm{com}}(\bm{q}_t)^2 \mathcal{F}_{\mathrm{com}}(\bm{R})^2}{\mathcal{F}_{\mathrm{com}}(\bm{R}-\bm{q}_t)\mathcal{F}_{\mathrm{com}}(\bm{R}+\bm{q}_t)}  \mathcal{F}_{\mathrm{rel}}(\bm{q}_t)^2.
	\label{eq:g2_final_vijays}
\end{equation}
In Fig.\ \ref{fig:doublewelltheory}, we compare Eq.\ (2) and Eq.\ (3), the density-noise correlation functions for the single-layer and bilayer systems.
There is a clear difference between the two models deep in the normal regime with correlation length $r_0$ much smaller than the system size used in the experiment, $\sim \SI{30}{\micro \meter}$. 
We use the numerical integration of Eq.~(2) (Eq.(3)) for the fitting of experimental data taken with single-layer (bilayer) systems \cite{Singh2014}.
The expression Eq.\ \eqref{eq:g2_final_vijays} holds in the presence of inter-layer tunnel-coupling as the relative and common modes fluctuate independently, both having Gaussian fluctuations for the experimentally relevant range of inter-layer coupling strengths \cite{Griifmmode2013}.

\subsection{Experimental procedure}

\subsubsection{Preparation of 2D Bose gases}
\begin{figure}[t]
	\includegraphics[width=0.75	\linewidth]{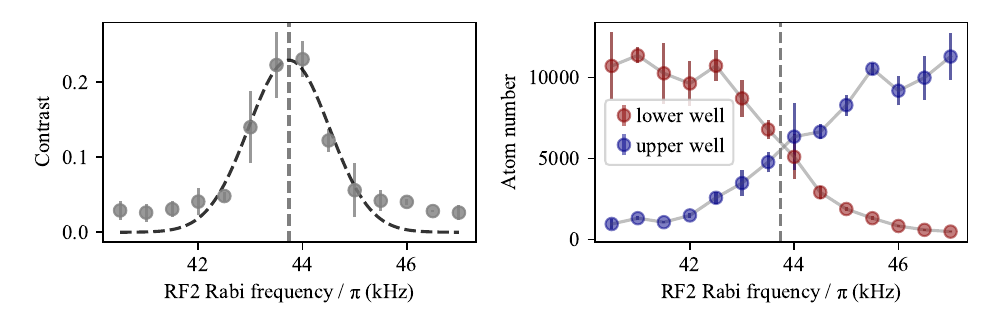}
	\caption{\label{fig:balance}
    Measurement of the double-well population imbalance and the matter-wave interference contrast.
    We vary the potential imbalance during the atom preparation process, by controlling the amplitude of one of three RFs field components used for generating the MRF-dressed double-well potential \cite{Beregi2024} (RF 2 amplitude, expressed in terms of Rabi frequency $\Omega = |g_F|\mu_B b/2\hbar$ where $g_F$ is the Lande g factor, $\mu_B$ is the Bohr magneton and $b$ is the AC magnetic field amplitude), while keeping the final potential shape the same.
    This enables good control of the population imbalance in the bilayer trap.
    (Left) the matter-wave interference contrast of the integrated column density distribution as a function of the RF field amplitude during the preparation process. 
    The dashed line is a Gaussian fit.
    The vertical dashed line is the frequency at the peak of the fitted Gaussian.
    Error bars correspond to a standard deviation of the observed values.
    The finite values of the contrast outside the peak region arises from imaging noise and fit errors: since the contrast is defined to be above zero, the fluctuations in the measured values result in finite positive values when averaged over samples, typically at $\sim 0.03$.
    (Right) The measured populations of the upper and lower potential wells as a function of the RF2 field amplitude. 
    The vertical dashed line is at the same location as in left panel, obtained from the fitted peak of the interference contrast.
    The line agrees well with the crossing of the two population traces indicating balanced bilayer populations.
        }
\end{figure}

We form the 2D potential for the laser-cooled \Rb{87} atoms using a combination of static and radiofrequency (RF) magnetic fields \cite{Harte2018}.
The static field is a quadrupole magnetic field with cylindrical symmetry about a vertical axis, and an additional single-frequency RF magnetic field is used for creating a single 2D Bose gas.
The RF frequency is 7.2 MHz with a quadrupole field gradient of 150 G/cm along $z$, resulting in the trap parameters described in the main text.
In contrast, three RF fields are applied to give a multiple-RF (MRF) double-well trap, at 7.08, 7.2 and 7.32 MHz, for the experiments to obtain data for Figs.~3 and 4.
Control over the amplitudes of RF components allows us to shape the double-well potential, as described in Refs.\ \cite{Harte2018,Bentine2020,Barker2020jphysb,Barker2020}, such  that it provides tight confinement in the vertical direction to produce a 2D potential.
We adiabatically load atoms into the MRF-dressed potential and let the system equilibrate for 500 ms, such that two quasicondensates are in equilibrium and independently fluctuating \cite{Sunami2022}.
As discussed in the main text, for the parameters used in this work, quasi-2D conditions  $\hbar \omega_z > k_B T$ and $\hbar \omega_z > \mu$ are satisfied: 
the presence of small excited state populations in the z direction at $\hbar \omega_z \simeq 1.2 k_B T$ results in a small reduction of the 2D interaction strength by $\lesssim$ 10\% however this changes the BKT critical point by less than 3\% as a result \cite{Holzmann2008, Fletcher2015}.
We ensure that the populations in the two wells are equal by maximizing the observed matter-wave interference contrast as described in~\cite{Barker2020}. 
To confirm this, we independently measure the atom populations of individual potential wells as well as the interference contrast, which are plotted in Fig.~\ref{fig:balance}. 
The peak interference contrast corresponds very closely to the crossing of the populations.

\begin{figure}[b]
	\includegraphics[width=0.45	\linewidth]{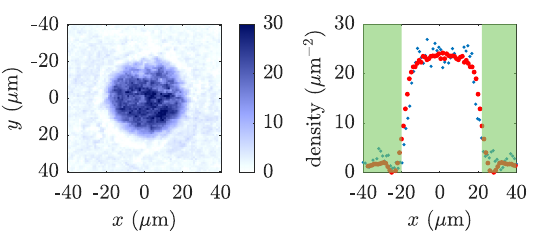}
	\caption{\label{fig:flatpot} Near-homogeneous density distribution realized with ring-shaped repulsive optical trap.
    (left) Single-shot image of in-situ density distribution, taken along the $z$ direction. 
    We apply repulsive optical dipole trap at 532 nm with ring-shaped geometry, generated by a DMD, onto trapped 2D Bose gases with weak harmonic confinement of $\omega_r/2\pi = \SI{8}{\hertz}$.
    (right) 1D plot of data in left panel, along $y=0$ axis (blue) as well as radially-averaged density distribution (red). 
    Green shaded region is where repulsive optical potential is applied.
		}
\end{figure}

For the experiments for Figs.\ 3 and 4, the additional box optical potential is created by far-detuned $\SI{532}{\nano \meter}$ laser shaped with a spatial light modulator (digital micromirror device, DMD) realizing a ring-shaped repulsive trap, on top of the weak harmonic potential of the MRF-dressed potential having $\omega_r/2\pi = \SI{8}{\hertz}$ (see Fig.\ \ref{fig:flatpot}).
While the resulting density is not completely uniform, the homogeneity is significantly better than for the harmonic potential thanks to increased density near the center of the trap.
We thus use the central region of the near-homogeneous system with radius of up to $\SI{15}{\micro \meter}$ for the data analysis.
The optical potential is slowly ramped up for a second after loading atoms into the RF-dressed potentials, such that the introduction of this additional potential does not cause excitations in the system.

\subsubsection{Alignment of the imaging system}
\begin{figure}[t]
	\includegraphics[width=0.45	\linewidth]{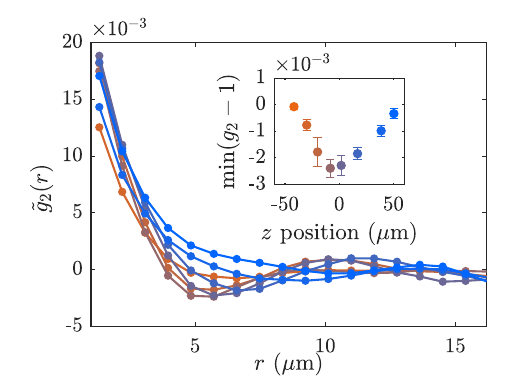}
	\caption{\label{fig:align} Focusing of the imaging apparatus by observation of the density-noise correlation functions.
    Correlation functions $\tilde{g}_2$, taken at $t=\SI{5.3}{\milli \second}$ at $\mathcal{D}=24$, are plotted for varying values of static magnetic bias field along $z$ direction, which shifts the position of the atoms from $\SI{-50}{\micro \meter}$ to $\SI{50}{\micro \meter}$. 
    We identify the optimal imaging configuration where the density pattern is most clearly resolved and amplitude of the anticorrelation peak of the $g_2$ is maximized (inset). 
		}
\end{figure}

Misalignment of the imaging system results in unintended artificial patterns in the obtained absorption images, affecting the density noise analysis \cite{Choi2012, Langen2013}.
Therefore, in addition to the selective imaging technique (see Fig.\ 1), we utilized the density noise correlation patterns as a focusing target with small-scale structure to perform precise alignment of the imaging system.
We have obtained the density-noise correlation functions at $t= \SI{5.3}{ms}$, as described in the text, with varying static magnetic bias field to precisely shift the location of the trapped cloud, and hence the cloud after the TOF expansion. We move the location of the sheet of repumping light accordingly, to move the imaged sample without moving the imaging apparatus including the objective lens.
In Fig.\ \ref{fig:align}, we show the density-noise correlation functions at varying relative position of the cloud after the TOF (with colors corresponding to the data points in the inset). 
We observe a clear change in the amplitude of the anticorrelation as a function of the location of the cloud, which is shown in inset as the change of minimum values of the correlation functions. 
The first minimum is deeper for sharper images, which indicates better focusing. 
We thus record the position of the atom after the TOF that minimizes the minimum value of the $\tilde{g}_2$ function, as the focal position of the imaging apparatus.

\subsubsection{Image analysis}
The analysis of the density noise patterns proceeds as follows. 
From at least 20 experimental images for each experimental parameter, we first normalize the images by the average density distribution for each dataset. 
We then obtain autocorrelation of images within a region of interest (ROI) which captures the central part of the cloud.
This results in a collection of correlation functions on a 2D grid, scaled by the squared mean density $n_0^2 = \langle \hat{n} (\bm{r}, t)\rangle ^2 = \langle \hat{\Psi}(\bm{r}, t)^\dagger \hat{\Psi}(\bm{r}, t)\rangle ^2$, where $ \hat{\Psi}(\bm{r}, t)$ is the bosonic field operator after the expansion, which corresponds to \cite{Singh2014}
\begin{equation}
	 \frac{ \langle\hat{n}(\bm{r}, t)\hat{n}(\bm{0}, t) \rangle }{n_0^2} = g_2(\bm{r}, t) + \frac{\delta(\bm{r})}{n_0},
\end{equation}
where the second term is the shot-noise term with zero mean, such that
\begin{equation}\label{eq:g2_def}
	g_2(\bm{r}, t) = \frac{ \langle \hat{\Psi}^{\dagger}(\bm{r}, t)\hat{\Psi}(\bm{r}, t) \hat{\Psi}^{\dagger}(\bm{0}, t)\hat{\Psi}(\bm{0}, t) \rangle }{\langle \hat{n}(\bm{r}, t) \rangle \langle \hat{n}(\bm{0}, t) \rangle },
\end{equation}
is identified by averaging over experimental repetitions.

\subsubsection{Effect of the finite thickness of the repumping region}
\begin{figure}[t]
	\includegraphics[width=0.6\linewidth]{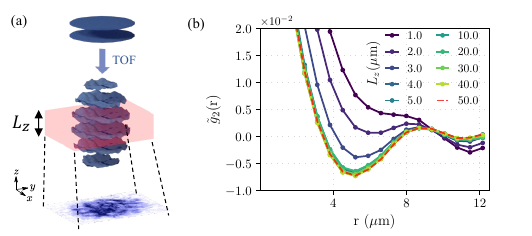}
	\caption{\label{fig:slice} Effect of integration of densities over a finite interval along $z$ axis for bilayer Bose gases.  
    (a) Finite thickness $L_z$ of the detection region results in the slicing of interference patterns along $z$ direction.
    (the wavelength of the interference fringes is exaggerated for illustration purposes). 
    (b) The simulated density-density correlation functions in the superfluid regime, for varying values of $L_z$.
    For this analysis, we simulated the 3D TOF expansion of bilayer systems with 3D time-dependent Gross-Pitaevskii solver, where the initial in-trap phase fluctuations are imprinted from the ones obtained from classical-field simulation~\cite{Sunami2022}.
    We integrated the obtained 3D density profiles with interference pattern along $z$ as illustrated in (a), over varying vertical range $L_z$ and obtained the 2D images.
    These simulated images are then analyzed using the same analysis pipeline used for experimental data shown in the main text to obtain the $g_2$ functions plotted.
		}
\end{figure}

The theoretical model in Sec.~\ref{sec:theory} assumes the full integration of the density pattern along the $z$ direction. 
However, in the experiment, we perform finite slicing of the density along the $z$ direction with thickness $L_z = \SI{10}{\micro \meter}$, to avoid the detrimental effect from the finite depth of field of the imaging system; 
for the case of bilayer systems, finite-thickness slicing may introduce additional effect from the relative-phase fluctuations through the slicing of interference patterns, as illustrated in Fig.~\ref{fig:slice} (a). 
To make a quantitative estimation of such an effect, we performed a numerical simulation of expansion dynamics by numerically solving the time-dependent 3D Gross-Pitaevskii equation with initial states being the bilayer-trapped 2D Bose gases with imprinted phase fluctuations from the classical-field simulations \cite{Sunami2022}.
Following the time evolution, we integrate the 3D density profiles along $z$ with varying integration range $L_z$ around the cloud center, corresponding to the density slicing with thickness $L_z$ in the experiment.
The integrated density noise patterns are then analyzed using the image analysis pipelines used for the experimental data shown in the main text, to obtain the $g_2(r)$ functions as plotted in Fig.~\ref{fig:slice} (b).
Different colors correspond to varying $L_{z}$.
As evident in Fig.~\ref{fig:slice} (b), for $L_{z} \gtrsim \SI{5}{\micro \meter}$, the correlation function $g_2$ is indistinguishable from the long-thickness limit that covers the entire cloud (red dashed line).

\subsection{Direct matter-wave interferometry readout of relative phase correlation function}

To observe the matter-wave interference, the trap is abruptly turned off, releasing the pair of 2D gases for a time-of-flight (TOF) expansion of duration $t = \SI{16.2}{ms}$.
Once released, a matter-wave interference pattern along $z$ appears.	
We image a thin slice of the density distribution with thickness $L_y = \SI{5}{\micro\metre}$, as described in detail in Ref.\ \cite{Sunami2022}.

We analyze the local phase of the matter-wave interference patterns following the procedure outlined in Refs. \cite{Barker2020,Sunami2022}, in which we perform a 1D fit of column density $\rho_x(z)$ at each location $x$ with 

\begin{equation}\label{eq:fringefit}
	\rho_x(z) = \rho_0 \exp\left(-z^2/2\sigma^2\right) \left[ 1 + c_0 \cos(kz+\theta(x)) \right],
\end{equation}
where $\rho_0,\sigma, c_0,k,\theta(x)$ are fit parameters. The extracted phase $\theta(x)$ encodes a specific realization of the fluctuations of the \textit{in situ} local relative phase between the pair of 2D gases \cite{Sunami2022}.
For each experimental run, we calculate the two-point phase correlation function $e^{i [\theta(x)-\theta(x')]}$ at locations $x$ and $x'$. 
We then determine the averaged correlation function
\begin{equation} \label{eq:corrmap}
	C^{(\mathrm{rel})} _{\mathrm{exp}}(x,x') = \frac{1}{N_r} \sum_j e^{i [\theta(x)-\theta(x')]},
\end{equation}
where the index $j$ runs over $N_r$ individual experimental realizations with $N_r>50$. 
We analyze the real part of the correlation function $C^{(\mathrm{rel})} (x,x')=\text{Re}\left[C^{(\mathrm{rel})} _{\mathrm{exp}}(x,x')\right]$, which is equal to $1$ for perfectly correlated pairs of points and $0$ for uncorrelated pairs of points.
$C^{(\mathrm{rel})}_{\mathrm{exp}} (x,x')$, is related to the relative phase correlation function 
\begin{equation}
	C^{(\mathrm{rel})} (\bm{r},\bm{r}'):= \frac{\langle \Psi_1^{\dagger}(\bm{r})\Psi_2(\bm{r})\Psi_2^{\dagger}(\bm{r}')\Psi_1(\bm{r}')\rangle}{\langle|\Psi_1(\bm{r})|^2\rangle\langle|\Psi_2(\bm{r}')|^2\rangle},
\end{equation}
For independently fluctuating layers, this is
\begin{equation}
	C^{(\mathrm{rel})} (\bm{r},\bm{r}') \simeq \frac{\langle \Psi^{\dagger}(\bm{r})\Psi(\bm{r'}) \rangle^2}{\langle |\Psi(\bm{r})|^2 \rangle \langle |\Psi(\bm{r'})|^2 \rangle} = \frac{g_1({\bm{r},\bm{r}'})^2}{n_0^2},
\end{equation}
where $n_0$ is the 2D density \cite{Sunami2022}. 
To quantify the decay of correlations, we calculate $C (\overline{x})$ by averaging $C^{(\text{rel})} (x,x')$ over points with the same spatial separation $\overline{x} = x-x'$ and perform fits with power-law models, obtaining the scaling of power-law exponent $\eta$ as shown in Fig.\ 4.

\twocolumngrid

\end{document}

%% file: refs.bbl
%

%% file: main.bbl
\begin{thebibliography}{45}
\bibitem [{tra()}]{trap_pos}%
  \BibitemOpen
  \href@noop {} {\bibinfo {title} {{We apply bias magnetic field to move the trapped atoms along $z$ direction while maintaining the same loading parameters, which allows the detection to be performed at the focus of the imaging apparatus for varying values of $t$ without reconfiguring the objective lens.}}}\BibitemShut {Stop}%
\bibitem [{int()}]{integrate_out}%
  \BibitemOpen
  \href@noop {} {\bibinfo {title} {{This is valid for the extent of integration along $z$ larger than a few repetitions of interference fringes with separation $\sim \SI{1}{\micro \meter}$.}}}\BibitemShut {Stop}%
\end{thebibliography}
